\documentstyle[11pt]{article}

\makeatletter
\def\citen#1{%
\edef\@tempa{\@ignspaftercomma,#1, \@end, }
\edef\@tempa{\expandafter\@ignendcommas\@tempa\@end}%
\if@filesw \immediate \write \@auxout {\string \citation {\@tempa}}\fi
\@tempcntb\m@ne \let\@h@ld\relax \def\@citea{}%
\@for \@citeb:=\@tempa\do {\@cmpresscites}%
\@h@ld}
\def\@ignspaftercomma#1, {\ifx\@end#1\@empty\else
   #1,\expandafter\@ignspaftercomma\fi}
\def\@ignendcommas,#1,\@end{#1}
\def\@cmpresscites{%
 \expandafter\let \expandafter\@B@citeB \csname b@\@citeb \endcsname
 \ifx\@B@citeB\relax 
    \@h@ld\@citea\@tempcntb\m@ne{\bf ?}%
    \@warning {Citation `\@citeb ' on page \thepage \space undefined}%
 \else
    \@tempcnta\@tempcntb \advance\@tempcnta\@ne
    \setbox\z@\hbox\bgroup 
    \ifnum0<0\@B@citeB \relax
       \egroup \@tempcntb\@B@citeB \relax
       \else \egroup \@tempcntb\m@ne \fi
    \ifnum\@tempcnta=\@tempcntb 
       \ifx\@h@ld\relax 
          \edef \@h@ld{\@citea\@B@citeB }%
       \else 
          \edef\@h@ld{\hbox{--}\penalty\@highpenalty
            \@B@citeB }%
       \fi
    \else   
       \@h@ld\@citea\@B@citeB
       \let\@h@ld\relax
 \fi\fi%
 \def\@citea{,\penalty\@highpenalty\hskip.13em plus.1em minus.1em}%
}
\def\@citex[#1]#2{\@cite{\citen{#2}}{#1}}%
\def\@cite#1#2{\leavevmode\unskip
  \ifnum\lastpenalty=\z@\penalty\@highpenalty\fi
  \ [{\multiply\@highpenalty 3 #1
      \if@tempswa,\penalty\@highpenalty\ #2\fi 
    }]\spacefactor\@m}
\makeatother

\let\a=\alpha

\def\nn{\nonumber} \def\bd{\begin{document}} \def\ed{\end{document}}
\def\ds{\documentstyle} \let\fr=\frac \let\bl=\bigl \let\br=\bigr
\let\Br=\Bigr \let\Bl=\Bigl 
\let\bm=\bibitem
\let\na=\nabla
\let\pa=\partial \let\ov=\overline 
\newcommand{\be}{\begin{equation}} 
\newcommand{\ee}{\end{equation}} 
\def\ba{\begin{array}}
\def\ea{\end{array}}
\def\ft#1#2{{\textstyle{{\scriptstyle #1}\over {\scriptstyle #2}}}}
\def\fft#1#2{{#1 \over #2}}
\def\del{\partial}
\def\vp{\varphi}
\def\sst#1{{\scriptscriptstyle #1}}
\def\oneone{\rlap 1\mkern4mu{\rm l}}
\def\simequiv{\buildrel\sim\over=}
\def\td{\tilde}
\def\wtd{\widetilde}
\def\ie{{\it i.e.\ }}
\def\im{{\rm i}}
\def\dalemb#1#2{{\vbox{\hrule height .#2pt
        \hbox{\vrule width.#2pt height#1pt \kern#1pt
                \vrule width.#2pt}
        \hrule height.#2pt}}}
\def\square{\mathord{\dalemb{6.8}{7}\hbox{\hskip1pt}}}
\def\R{\rlap{\rm I}\mkern3mu{\rm R}}
\def\E{\rlap{\rm I}\mkern3mu{\rm E}}
\def\Z{\rlap{\sf Z}\mkern3mu{\sf Z}}
\newcommand{\ho}[1]{$\, ^{#1}$}
\newcommand{\hoch}[1]{$\, ^{#1}$}
\newcommand{\bea}{\begin{eqnarray}} 
\newcommand{\eea}{\end{eqnarray}} 
\newcommand{\ra}{\rightarrow}
\newcommand{\lra}{\longrightarrow}
\newcommand{\Lra}{\Leftrightarrow}
\newcommand{\ap}{\alpha^\prime}
\newcommand{\bp}{\tilde \beta^\prime}
\newcommand{\tr}{{\rm tr} }
\newcommand{\Tr}{{\rm Tr} } 
\newcommand{\NP}{Nucl. Phys. }

\newcommand{\sissatamphys}{\it SISSA, Via Beirut No. 2-4, 34013 Trieste, 
Italy and\\
Center for Theoretical Physics,
Texas A\&M University, College Station, Texas 77843}
\newcommand{\ens}{\it Laboratoire de Physique Th\'eorique de l'\'Ecole
Normale Sup\'erieure\hoch{2}\\
24 Rue Lhomond - 75231 Paris CEDEX 05}
\newcommand{\ic}{\it The Blackett Laboratory, Imperial College,\\
Prince Consort Road, London SW7 2BZ, UK}

\newcommand{\auth}{E. Cremmer\hoch{\dagger},  
H. L\"u\hoch{\dagger}, C.N. Pope\hoch{\ddagger1} and K.S. Stelle\hoch{\star}}

\begin{document}
\begin{flushright}
\hfill{CTP TAMU-29/97}\\
\hfill{Imperial/TP/96-97/54}\\
\hfill{LPTENS-97/34}\\
\hfill{SISSARef.\ 92/97/EP}\\
\hfill{hep-th/9707207}\\
\hfill{July 1997}\\
\end{flushright}

\begin{center}
{\bf{\large Spectrum-generating Symmetries for BPS
Solitons}\hoch{\diamondsuit}}

\vspace{15pt}
\auth

\vspace{10pt}

{\hoch{\dagger}\ens}

\vspace{10pt}
{\hoch{\ddagger}\sissatamphys}

\vspace{10pt}
{\hoch{\star}\ic}

\vspace{15pt}

\underline{ABSTRACT}
\end{center}

     We show that there exist nonlinearly realised duality symmetries
that are independent of the standard supergravity global symmetries,
and which provide active spectrum-generating symmetries for the
fundamental BPS solitons.  The additional ingredient, in any spacetime
dimension, is a single scaling transformation that allows one to map
between BPS solitons with different masses.  Without the inclusion of
this additional transformation, which is a symmetry of the classical
equations of motion, but not the action, it is not possible to find a
spectrum-generating symmetry.  The necessity of including this scaling
transformation highlights the vulnerability of duality multiplets to
quantum anomalies.  We argue that fundamental BPS solitons may be
immune to this threat.

{\vfill\leftline{}\vfill
\vskip	10pt
\footnoterule
{\footnotesize \hoch{\diamondsuit} Research supported in part by the
European Commission under TMR contract\\
\phantom{EEEee}ERBFMRX-CT96-0045 \vskip -12pt} \vskip 10pt
{\footnotesize	\hoch{1} Research supported in part by DOE 
Grant DE-FG03-95ER40917 and the \\
\phantom{EEee\, } EC Human Capital and Mobility Programme
under contract ERBCHBGT920176.\vskip	-12pt} \vskip 10pt
{\footnotesize
        \hoch{2} Unit\'e Propre du Centre National de la Recherche
Scientifique, associ\'ee \`a l'\'Ecole Normale Sup\'erieure\\
\phantom{EEEee}et \`a
l'Universit\'e de Paris-Sud}}

\pagebreak
\setcounter{page}{1}

\section{Introduction}

     The study of BPS-saturated solitons in the supergravities that
describe the low-energy limits of string theories has proved to be a
valuable tool for elucidating the non-perturbative structures of these
theories.  These solitons are solutions in which infinite
$p$-branes occupy a longitudinal submanifold in spacetime, with the
fields depending on the coordinates of the transverse space.  These
fields include one or more of the antisymmetric tensor field strengths
in the supergravity theory, which carry either electric or magnetic
type charges (or both, in the case of dyonic $p$-branes).  In fact,
these solutions are characterised by the configuration of
non-vanishing charges, and by the asymptotic values of the scalar
fields at infinity.  These asymptotic values can be thought of as the
modulus parameters for the solution.  If we restrict attention to
BPS-saturated solutions, then the mass is not an independent
parameter, but is instead some function of the charges and the scalar
moduli.  It is useful to try to organise the various solutions into
multiplets, by making use of the global symmetries of the supergravity
theory.  

     There indeed exist continuous global symmetries in supergravity
theories \cite{cj1,cj2}, which act linearly on the charges,
nonlinearly on the scalars, and which leave the Einstein-frame metric
invariant. In the following, we shall refer to these symmetries as the
standard supergravity global symmetries. The orbits of these standard
global symmetry groups $G$ yield families of $p$-brane solutions.  In
particular, the maximal compact subgroup $H$ of $G$ is the stability
group of a point in modulus space; this allows the set of charges to
be rotated while holding the asymptotic values of the scalars fixed.
This rotation in the vector space of the charges holds an invariant
quadratic expression in the charge vectors fixed, whose square root
may be thought of as an $H$-invariant ``length.''  Such a rotation,
and indeed any transformation under the standard global symmetry group
$G$, preserves the mass of the solution. Put another way, the
expression for the mass as a function of the charges and the scalar
moduli is invariant under $G$.  This can be immediately seen from the
fact that $G$ leaves the Einstein-frame metric invariant.  The
implication of these observations is that the standard global
supergravity symmetry groups are insufficient for the purpose of
generating complete sets of $p$-brane solitons.

     The problem of finding a solution-generating symmetry arises in a
more severe form at the quantum level in string theory.  It has been
argued that in the quantised string theory only a discrete subgroup
of the classical supergravity symmetry $G$ can be consistent with the
Dirac quantisation condition, which allows only a discrete lattice of
charges for any given vacuum.\footnote{Actually, as we shall argue below,
it is really the spectrum-generating symmetry, and not the standard
supergravity symmetry $G$, that is discretised by virtue of the Dirac
quantisation condition.}  The discrete subgroup $G(\Z)$
is known as U-duality, and is conjectured to survive as an exact
symmetry even at the non-perturbative level \cite{ht}.  In fact, in
string theory it has been proposed that states should be {\it
identified} under U-duality.  Thus, not only does the U-duality
group $G(\Z)$ suffer from the same deficiency as the classical group
$G$ for generating independent solutions, but also those solutions
that it can relate are taken to be identical, and so the $G(\Z)$ orbit
consists of one single state.  This is not what one could call a
satisfactory spectrum-generating symmetry.

    Despite these shortcomings, there is a certain sense in which the
orbit of the U-duality group is associated with the spectrum of
distinct BPS quantum states.  If one looks only at the action of the
U-duality group on the charge lattice, and ignores its action on the
scalar moduli, then it does map between allowed charge vectors.  If
these charge vectors were taken to be associated with a single fixed
vacuum, then one would indeed have the spectrum of physically-distinct
states.  In Ref.\ \cite{ht}, a procedure of ``analytic continuation in
the moduli'' was proposed, to return the moduli after a U-duality 
transformation to their initial values.  This procedure, however, does
not make clear whether there is an actual symmetry transformation in
the theory that can implement this analytic continuation, and so it
does not clearly give a proper spectrum-generating symmetry. 

   In this paper, we shall show that there exists a different $G$
symmetry (although with the same abstract group $G$), realised
nonlinearly on the fields of the theory, that holds the scalar moduli
fixed while transforming the charge vectors in a linear fashion.  This
is the true spectrum-generating group, which we shall call the {\it
active} $G$ symmetry group.  The key point is to recognise that the
actual classical global symmetry group in any supergravity theory is
larger than is customarily presented, and includes an additional
scaling transformation, which is a symmetry of the equations of motion
corresponding to a homogeneous scaling of the action.  A well-known
example of such a symmetry is in pure Einstein gravity, where there is
a global scaling symmetry of the equations of motion under the
transformation $g_{\mu\nu}\rightarrow \lambda^2\, g_{\mu\nu}$.  This
symmetry can be used to explain the organisation of solutions into
one-parameter families, such as the Schwarzschild solution, where the
mass is a free parameter.  Analogous scaling symmetries exist in all
supergravity theories.  Because they allow one to scale magnitudes in
and out, we shall call these scaling symmetries ``trombone''
symmetries.  At the classical level, taken together with the standard
supergravity $G$ symmetry,\footnote{Note that at least at the
classical level the trombone symmetry is as much a valid symmetry of
the theory as the standard global symmetry $G$. In fact in even
dimensions, the latter shares with the trombone symmetry the feature
that it is a symmetry only of the equations of motion, but not of the
action.} they allow one to reach the entire parameter space of BPS
solitons that preserve half the supersymmetry.  Thus, the entire
classical family of such fundamental BPS solutions can be reached by
the application of symmetries of the theory.  More precisely, it is
the stability group $H$ of the moduli together with the trombone
symmetry, that allow one to reach arbitrary points in the
charge-vector space while holding the scalar moduli fixed.

   At the quantum level, the discussion becomes more involved.  One
might think that one could simply take the direct product of the
trombone symmetry with the $H$ subgroup of the standard $G$ symmetry,
and then make a restriction to a discrete linearly-realised subgroup of
this product that is compatible with the charge lattice required by the
Dirac quantisation condition.  However, there is no such group. The
reason for this is that the trombone scaling symmetry and the $H$
symmetry would need to be independently restricted to discrete
subgroups, and there is in general no way to do this for either factor
in such a way as to ensure that only allowed charge-lattice points can
be related.  What one must do instead is to construct a nonlinear
realisation of the full $G(\Z)$ group, linearly realised on $H$, using
the trombone symmetry as a compensating transformation. 

     It is worth emphasising that the Dirac quantisation condition {\it
by itself} requires that the classical spectrum-generating symmetry of a
supergravity theory be restricted to a discrete $G(\Z)$ subgroup.  This
is quite different from the situation for the standard supergravity
global symmetry groups, since the latter move not only the charges but
also the scalar moduli, and the Dirac quantisation condition {\it by
itself} does not require that the charge lattices for different points
in the modulus space must coincide (even though the lattice at each
modulus point must respect the quantisation condition).  Thus the
discretisation of the U-duality groups in \cite{ht} must arise for reasons
that go beyond the Dirac quantisation condition. In fact it has been argued
that since the T-duality groups are subgroups of the U-duality groups, then
the discretisation of the former (based on perturbative string-theoretic
considerations) provides supporting evidence for the discretisation of the 
latter \cite{ht}.

     We shall be concerned with $p$-brane solitons in
$D=11$ supergravity, type IIB supergravity, and their dimensional
reductions.  All of these theories have analogous scaling symmetries;
in fact, in the dimensionally-reduced theories they are a direct
consequence of the corresponding symmetries of the higher-dimensional
$D=11$ or type IIB theories.  In $D=11$, the bosonic Lagrangian is \cite{cjs}
\be
{\cal L}=eR -\ft1{48}e\, F_4^2 +\ft16 \ast(dA_3\wedge dA_3\wedge A_3)\ ,
\label{d11lag}
\ee
where $F_4=dA_3$, $A_3$ is the 3-form potential, and $e$ is the determinant of
the vielbein. The corresponding trombone symmetry is
\be
g_{\mu\nu}\longrightarrow \lambda^2\, g_{\mu\nu}\ ,\qquad 
A_3\longrightarrow \lambda^3\, A_3\ ,\label{d11trombone}
\ee
under which the Lagrangian scales as ${\cal L}\rightarrow \lambda^9\,
{\cal L}$. The equations of motion scale homogeneously, and thus
(\ref{d11trombone}) is a symmetry of the equations of motion.  This
symmetry is responsible for the
extremal membrane and 5-brane classical soliton solutions occurring in
one-parameter families with
arbitrary values of the charge.  The symmetry (\ref{d11trombone}) is
preserved under Kaluza-Klein dimensional reduction, after which the
metric still scales as $g_{\mu\nu}\rightarrow \lambda^2\, g_{\mu\nu}$,
and in addition all $n$-index potentials scale with a factor
$\lambda^n$, while all scalar fields are left invariant.  (Trombone
symmetries were also used in the rheonomy approach to supersymmetry;
see, for example \cite{cdf}.) The same
rescaling rules apply to the trombone symmetry of the type IIB
supergravity theory. We shall see that this rather humble scaling
symmetry plays a central r\^ole in permitting the construction of
solution-generating symmetry transformations that map actively between
physically-inequivalent soliton solutions.
     
     The simplest non-trivial example of a solution-generating global
symmetry is when $G=SL(2,\R)$.  We shall accordingly consider in
detail the example of string solitons in the type IIB supergravity
theory, for which $SL(2,\R)$ is the symmetry group.  (Another example
would be provided by S-duality in the $D=4$ heterotic string theory
\cite{sen}.)  The layout of the rest of this paper is as follows.  In
section 2, we describe the type IIB supergravity theory and the action
of the standard $G=SL(2,\R)$ symmetry on its fields. In section 3, we
construct the nonlinearly realised active $SL(2,\Z)$ symmetry, and we
discuss its group structure in section 4. In section 5, we give a
group-theoretical interpretation of the charge spectrum, using some
elements of the earlier construction.  In section 6 we generalise the
discussion to the lower-dimensional cases with larger symmetry groups,
and in section 7 we consider the problem of quantum anomalies in the
spectrum-generating symmetry.  We end with a conclusion in section 8.

\section{Type IIB in $D=10$}

     The low-energy effective theory for the type IIB string is type
IIB supergravity, whose bosonic fields comprise the metric, a dilaton
$\phi$, an axion $\chi$, two 2-form potentials $A_2^{(i)}$, and a
4-form potential whose associated field strength is self dual.  The
4-form potential, $\chi$, and $A_2^{(2)}$ are R-R fields, and the
remainder are NS-NS.  Owing to the self-duality of the 5-form field
strength, there is no simple way to write a covariant Lagrangian for
these fields alone.  However, by adding extra degrees of freedom,
namely by removing the self-duality condition, one can write a
Lagrangian whose equations of motion yield the type IIB equations after
imposing by hand the self-duality constraint as a consistent
truncation \cite{bbo}.  Thus, our starting point is the Lagrangian
\bea
{\cal L} &=& eR +\ft14 e\, {\rm tr}(\del_\mu{\cal M}^{-1}\, \del^\mu{\cal M})
-\ft1{12}e\, H_3^T\, {\cal M}\, H_3 -\ft1{240} e\, H_5^2 \nn\\
&&- \ft1{2\sqrt2} \epsilon_{ij}\,\ast(B_4\wedge 
dA_2^{(i)}\wedge dA_2^{(j)})\ ,\qquad \label{2blagm}\\
&=& eR -\ft12 e\, (\del\phi)^2 -\ft12 e\, e^{2\phi}\, (\del\chi)^2 
-\ft1{12} e\, e^{-\phi}\, (F_3^{(1)})^2 -\ft1{12} e\, e^\phi\, 
(F_3^{(2)})^2 \nn\\
&& -\ft1{240} e\, H_5^2 - \ft1{2\sqrt2} \epsilon_{ij}\,\ast(B_4\wedge 
dA_2^{(i)}\wedge dA_2^{(j)})\ ,\label{2blag}
\eea
where 
\be
{\cal M}= \pmatrix{ e^{-\phi} + \chi^2\, e^\phi & \chi\, e^{\phi} \cr
                    \chi\, e^\phi & e^\phi }\ ,\qquad\qquad
H_3 =\pmatrix{dA_2^{(1)}\cr dA_2^{(2)} }\ .\label{mato}
\ee
The field strengths appearing in (\ref{2blag}) are defined as follows:
\be
F_3^{(1)}=dA_2^{(1)}\ , \qquad F_3^{(2)} = dA_2^{(2)} +\chi\, dA_2^{(1)}\ ,
\qquad H_5 =dB_4 + \ft1{2\sqrt2} \epsilon_{ij} A_2^{(i)}\wedge dA_2^{(j)}
\ .\label{2bcs}
\ee

     The equations of motion following from (\ref{2blag}) are
\bea
R_{\mu\nu} &=& \ft12 \del_\mu\phi\, \del_\nu\phi +\ft12 e^{2\phi}\, 
\del_\mu\chi\, \del_\nu\chi + \ft1{48}(H^2_{\mu\nu} -\ft1{10}
 H_5^2\, g_{\mu\nu})\label{ein} \\
&&+ \ft14e^{-\phi} ((F^{(1)})^2_{\mu\nu} -\ft1{12} (F_3^{(1)})^2 g_{\mu\nu})
\nn\\
&&+\ft14e^{\phi} ((F^{(2)})^2_{\mu\nu} -\ft1{12} (F_3^{(2)})^2 g_{\mu\nu})\ ,
\nn\\
\nabla^\mu\, H_{\mu\nu\rho\sigma\lambda} &=& \ft{1}{72\sqrt2}\, 
\epsilon_{ij}\,
\epsilon_{\nu\rho\sigma\lambda}{}^{\mu_1\cdots\mu_6}\, 
F_{\mu_1\mu_2\mu_3}^{(i)}\, F_{\mu_4\mu_5\mu_6}^{(j)} \ ,\label{5feq}\\
\nabla^\mu\, (e^{-\phi} \, F_{\mu\nu\rho}^{(1)} -e^\phi\, \chi \,
F_{\mu\nu\rho}^{(2)}) &=& -\ft{1}{6\sqrt2}\, 
H_{\nu\rho}{}^{\mu\lambda\sigma} \,
(F_{\mu\lambda\sigma}^{(2)} +\chi\, F_{\mu\lambda\sigma}^{(1)})\ ,\label{h1}\\
\nabla^\mu\,( e^{\phi} \, F_{\mu\nu\rho}^{(2)})&=& \ft{1}{6\sqrt2}\, 
H_{\nu\rho}{}^{\mu\lambda\sigma} \,F_{\mu\lambda\sigma}^{(1)}\ ,\label{h2}\\
\nabla^\mu(e^{2\phi}\, \del_\mu \chi) &=&\ft16\, e^\phi\,
F_{\mu\nu\rho}^{(1)}\,  F^{(2)\mu\nu\rho}\ ,
\label{chieq} \\
\square\phi &=&  e^{2\phi}\, (\del\chi)^2 +\ft1{12}\, e^{-\phi}\,
(F_3^{(1)})^2 -\ft1{12}\, e^{\phi}\, (F_3^{(2)})^2\ ,\label{phieq}
\eea
Note that the equation for $H_5$ can be rewritten as $d\ast H_5=
\ft1{2\sqrt2}\,\epsilon_{ij}\, F_3^{(i)}\wedge  F_3^{(j)}$.  Since we also 
have the Bianchi
identity $dH_5=\ft1{2\sqrt2}\,\epsilon_{ij}\, F_3^{(i)}\wedge 
F_3^{(j)} $, we see that we can
consistently impose the self-duality condition $H_5=\ast H_5$.  After doing
this, the equations (\ref{ein}-\ref{phieq}) become precisely the field
equations of type IIB supergravity \cite{IIB}. 
The Lagrangian (\ref{2blag}) is manifestly
$SL(2,\R)$ invariant.

     The action of $SL(2,\R)$ on the fields can be expressed as 
\be
H_3\longrightarrow (\Lambda^T)^{-1}\, H_3\ ,\qquad 
{\cal M}\longrightarrow \Lambda\, {\cal M}\, \Lambda^T\ ,\label{tran}
\ee
where 
\be
\Lambda = \pmatrix{a & b \cr c & d}\ ,\label{mat}
\ee
and $ad-bc =1$.  Defining the complex scalar field $\tau=\chi + i\, 
e^{-\phi}$, the transformation on ${\cal M}$ can be seen to imply that
\be
\tau \longrightarrow \fft{a \tau + b}{c\tau + d}\ .
\ee
Note that since $H_5$ is a singlet under $SL(2,\R)$, the self-duality 
constraint which is imposed by hand also preserves 
the $SL(2,\R)$ 
symmetry.

In addition to the $SL(2,\R)$ symmetry described above, there is also a 
trombone scaling symmetry, as we anticipated in the Introduction:
\be
g_{\mu\nu}\longrightarrow \lambda^2\, g_{\mu\nu}\ ,\qquad
A_2^{(i)}\longrightarrow \lambda^2\, A_2^{(i)}\ ,\qquad
H_5\longrightarrow \lambda^4\, H_5\ .
\ee
It is important to note that this rescaling leaves the scalar fields
$\phi$ and $\chi$ invariant.  This is also true of the scalar fields
of all the lower-dimensional supergravities that we shall consider. Thus
the full global symmetry group of type IIB supergravity is $GL(2,\R)$.

     We shall first consider classical string soliton solutions.
These are characterised by two electric charges $Q_e=(p,q)$ (carried
by the NS-NS and R-R 3-form field strengths), and by the two scalar
moduli $\phi_0$ and $\chi_0$, corresponding to the asymptotic values
at infinity of the dilaton $\phi$ and the axion $\chi$.  The string
coupling constant is given by $g=e^{\phi_0}$.  The full
4-parameter family of solutions can be generated starting from a
specific solution, for example from a pure NS-NS string with vanishing
moduli $\phi_0$ and $\chi_0$, by acting with the $SL(2,\R)$ and
trombone symmetries of the equations of motion.  

     The action of the $SL(2,\R)$ symmetry on the parameters of the
solutions is
\be
Q_e\equiv \pmatrix{p \cr q} \longrightarrow \Lambda\, 
\pmatrix{p \cr q}\ ,
\qquad \tau_0\longrightarrow \fft{a\, \tau_0 + b}{c\, \tau_0 + d}\ .
\label{sl2r}
\ee
The $SL(2,\R)$ transformation rule for the electric charges follows from the 
fact that the field equations (\ref{h1}) and (\ref{h2}) for the two 
3-form field strengths can be rewritten as 
\be
d\ast ({\cal M}\, H_3) = -\ft1{\sqrt2}\, H_5\wedge \Omega H_3\ ,
\qquad\qquad \Omega =
\pmatrix{0 & 1\cr -1 & 0}\ ,\label{omega}
\ee
and so the canonical electric Noether charges are given by
\be
Q_e=\pmatrix{p\cr q} = \int\Big(\ast {\cal M} H_3 + \ft1{3\sqrt2}
\, \Omega (2B_4\wedge H_3-H_5\wedge A_2)\Big)\ .
\ee
(This expression is in fact invariant under the gauge transformations
$\delta A_2^{(i)}= d\Lambda_1^{(i)}$, despite the appearance of bare
2-form potentials, since the gauge invariance of $H_5$ requires a
compensating transformation $\delta B_4 = -\ft1{2\sqrt2}\epsilon_{ij}\,
\Lambda_1^{(i)} \wedge dA_2^{(j)}$, as can be seen from (\ref{2bcs}).) 
From the transformation rules (\ref{tran}) for ${\cal M}$ and $H_3$, and
from the invariance $\Omega =\Lambda\Omega\Lambda^T$ of $\Omega$,
defined in (\ref{omega}), the $SL(2,\R)$ transformation rule given in
(\ref{sl2r}) for the electric charges follows. Note that the magnetic
charges, by contrast, are defined by
\be
Q_m=\fft1{2\pi}\int H_3\ ,
\ee 
and so they transform as $Q_m\longrightarrow 
(\Lambda^T)^{-1}\, Q_m$ under $SL(2,\R)$.  (This is indeed consistent with 
the fact that the Dirac quantisation condition
is $Q_m^T\, Q_e =$ integer, and must be preserved under the
standard $SL(2,\R)$.) In the rest of this paper, we shall be considering
only electric charges, and shall therefore drop the subscript
`$\scriptstyle e$' on $Q$.

     The action of the trombone symmetry on the charge and modulus
parameters is simply incorporated by relaxing the condition $ad-bc=1$;
in other words, the full symmetry has the group structure $GL(2,\R)$.
This indeed has four parameters, consistent with the fact that an
arbitrary BPS string solution can be obtained from any specific one by
a $GL(2,\R)$ transformation.

     At the quantum level, due account must be taken of the Dirac
quantisation conditions between electrically-charged strings and
magnetically-charged 5-branes.  These conditions imply that the
electric charges $Q$ in a given fixed scalar vacuum should lie on a
lattice, of which the simplest choice is to take $Q=(p,q)$, with $p$
and $q$ integers.  (An equally valid choice would be the lattice
obtained from this by acting on the charge vectors with an arbitrary
fixed $SL(2,\R)$ transformation with the magnetic charge lattice
transforming in the appropriate contragedient fashion.)  If the
standard global $SL(2,\R)$ symmetry had acted only on the charges, and
not also on the scalar moduli, it would be discretised by the Dirac
quantisation condition since only an $SL(2,\Z)$ subgroup would
preserve the chosen charge lattice.  Since the standard $SL(2,\R)$
actually moves the scalar moduli at the same time as moving the
charges, the Dirac quantisation condition by itself does not imply
that this group must be discretised.  In fact, as we shall show in the
next section, there is a nonlinearly realised active $SL(2,\R)$
symmetry that does move just the charges, while holding the moduli
fixed, and it is the discretisation to $SL(2,\Z)$ of this symmetry
that can be deduced from the Dirac quantisation condition.  In the
simple case where the electric charges are integers, $Q=(p,q)$, this
active $SL(2,\Z)$ is defined by the requirement that the entries $a$,
$b$, $c$ and $d$ be integers. These active $SL(2,\Z)$ transformations
can generate the entire charge lattice starting just from the set of
charges $Q=(n,0)$. Starting from $Q=(1,0)$, and acting with an active
$SL(2,\Z)$ transformation of the form (\ref{mat}), but with $a$, $b$,
$c$ and $d$ now integers satisfying $a d - b c =1$, one may verify
that one generates integer pairs $(p,q)$ with $p$ and $q$ relatively
prime. Starting instead from $(n,0)$ gives the set $(p,q)$ for all
integer pairs with a common factor $n$. Since solitonic $p$-brane
solutions with charges $(n,0)$ can be viewed as coincident
superpositions of $n$ $p$-branes with charge $(1,0)$, the spectrum of
elementary single $p$-branes will be taken to be the irreducible
active $SL(2,\Z)$ orbit of solutions containing the $(1,0)$
solution. We shall call this the elementary $SL(2,\Z)$ orbit.

\section{Construction of the active $SL(2,\Z)$}

     The standard supergravity $SL(2,\R)$ symmetry has the property of
transforming the charges and the scalar moduli at the same time. In
string theory, states that are related by an $SL(2,\Z)$ subgroup of
this standard $SL(2,\R)$ symmetry are treated as equivalent.  In other
words, this $SL(2,\Z)$ is interpreted as a local or ``gauged''
symmetry. In particular, this implies that the full charge lattice of
solutions in a given vacuum $(\phi_0,\chi_0)$ is identified with the
full charge lattice of solutions for any other vacuum whose scalar
moduli are related to $(\phi_0,\chi_0)$ by a gauged $SL(2,\Z)$
transformation.  However, in string theory, the physical spectrum of
distinct string solitons is described by the full charge lattice of
solutions for a given {\it fixed} vacuum, {\it i.e.}\ with
$(\phi_0,\chi_0)$ fixed.  Clearly, the $SL(2,\Z)$ subgroup of the
standard $SL(2,\R)$ cannot generate this multiplet of distinct charge
states in a given vacuum.  In particular, at fixed $(\phi_0,\chi_0)$,
the masses of the various strings at different charge-lattice points
will in general be different, and so it is evident that no subgroup of
the standard classical $SL(2,\R)$ can possibly relate them.  In fact,
the extra ingredient that is needed in order to construct the
multiplets of physically-distinct solutions is the trombone symmetry,
which does rescale the masses.

     Classically, it is easy to see that the trombone symmetry,
together with the $H=SO(2)$ maximal compact subgroup of $G=SL(2,\R)$,
can be used to generate a solution with arbitrary charges $Q=(p,q)$,
while holding $(\phi_0,\chi_0)$ fixed.  This is because, as we
remarked previously, the $H$ subgroup rotates the charge vector while
keeping its invariant length fixed.  Combined with the rescaling
trombone symmetry, which also preserves the scalar moduli, the entire
plane of charges can be reached.  (In fact the $H$ subgroup, together
with a rescaling of the charges, was used at the level of string
solutions in \cite{schwarz} to obtain the general family of $Q=(p,q)$
string solitons for arbitrary fixed scalar moduli in the type IIB
theory.)

    At the quantum level, one might be tempted to try to take the
direct product of the $H$ subgroup times the trombone scaling
symmetry, acting linearly on the charge-vector space, and then to
search for a subgroup of this product that maps only between the
points on the charge lattice which are allowed by the Dirac
quantisation condition. Unfortunately, there is no subgroup that does
this. For the trombone symmetry, this can be seen because this
linearly-realised transformation acts multiplicatively on the charges,
and only by multiplying by the integers could one ensure that only
maps between lattice points occur. However, the integers with
multiplication taken as the composition operation do not form a group,
except for a $Z_2$ generated by $-1$, which leaves the mass invariant.
This is clearly insufficient for the purposes of generating the
spectrum.  Similarly, the $H=SO(2)$ stability group does not in
general admit a discrete subgroup that preserves the given charge
lattice.  An exceptional situation, where a lattice-preserving
discrete subgroup of $H$ does occur, arises for a specific point in
the modulus space, and for those points related to it by $SL(2,\Z)$
transformations that preserve the given charge lattice. For these
special moduli there is a lattice-preserving $Z_2$ subgroup of $H$.
For example, for the case of the integer charge lattice, the special
modulus point is the self-dual value $\tau_0=i$, for which
$\chi_0=\phi_0=0$.  In fact this $Z_2$ subgroup is the Weyl group of
$SL(2,\R)$, and in lower dimensions, where the global symmetry groups
$G$ are larger, the lattice-preserving subgroup of $H$ for appropriate
special values of the scalar moduli again turns out to be the Weyl
group of $G$ \cite{weyl,vertical}.  In any case, however, the Weyl
group is not sufficiently large to allow the whole charge lattice to
be generated, even at these special modulus values.  This is because
the action of the Weyl group is to permute the axes of the
charge-vector space, but it does not allow any intermediate rotations.

    The way to solve this quantum-level problem is to change over to a
nonlinear realisation of $SL(2,\Z)$, acting on the fields of the
theory through $H=SO(2)$ together with compensating trombone
transformations, in a way which happens to coincide with the
linearly-realised $SL(2,\Z)$ when acting on the charge vectors, but
leaving the vacuum modulus fixed. Since this symmetry coincides with
the known action of $SL(2,\Z)$ on the charge vectors, it clearly
preserves the charge lattice.

    Let us begin by trying to make use of the trombone symmetry to
generate the multiplet of quantum-level $(p,q)$ strings in a fixed
vacuum. The idea will be to perform an $SL(2,\Z)$ transformation on a
given charge pair, for example $Q=(1,0)$, and then to follow this
with a compensating transformation that preserves the new charges, but
restores the transformed scalar moduli back to their original values.
This compensating transformation can in fact be decomposed as a
product of two factors.  First, there is a certain subgroup of the
$SL(2,\R)$ symmetry group, obtained by conjugating its Borel subgroup
by the denominator group $H=SO(2)$, that preserves the charge vector
up to an overall scale,\footnote{This property of the Borel subgroup
of $SL(2,\R)$ also plays an important role in the twistor-based
formulation of a superparticle in $d=2+1$ dimensions; if one views
$SL(2,\R)$ as ${\it Spin}(2,1)$, this construction also naturally
generalises to Lorentz groups in higher dimensions, treating the
Lorentz group ${\it Spin}(d-1,1)$ as the conformal group of the sphere
${\cal S}^{d-2}$ \cite{lorentz}.} while allowing the scalar moduli to
be transformed to arbitrary values.  We can use this subgroup to
restore the scalar moduli to their original values.  Secondly, we can
apply a trombone rescaling to restore the rotated charges to their
proper normalisations, while leaving the scalar moduli fixed at their
now-restored values.  The group structure of the combined
Borel-trombone transformations, a subgroup of $GL(2,\R)$, is
isomorphic to that of the Borel subgroup of $SL(2,\R)$. The net effect
of the $SL(2,\Z)$ transformation followed by the Borel-trombone
compensator is to give a solution at a new point on the charge
lattice, while remaining in the original vacuum.  The full set of such
compensated $SL(2,\Z)$ transformations will generate the entire
irreducible $SL(2,\Z)$ multiplet of $(p,q)$ strings in a given vacuum.
Note that the compensator transformations do not need to be
integer-valued, since they leave the charges fixed.

     To be specific, consider the case of a single NS-NS string, with
electric charges $Q=(p,q)$. For simplicity, begin with the scalar
vacuum defined by $\phi_0=0$, $\chi_0=0$, {\it i.e.}\ at the self-dual
point $\tau_0=i$. The Dirac quantization condition can be satisfied by
choosing to restrict $Q$ to be of the form $Q=(p,q)$, where $p$
and $q$ are arbitrary integers.  In order for this to lie on the
elementary $SL(2,\Z)$ orbit we must take $p$ and $q$ to be relatively
prime. Application of $SL(2,\Z)$ transformations will then map
transitively around the irreducible lattice of elementary charge
states. Starting from $Q_1=(p_1,q_1)^{\rm T}$, one thus arrives at the
charge
\be
Q_2= \pmatrix{p_2\cr q_2} =\Lambda Q_1\ .\label{LQ}
\ee
In the process, however, the scalar moduli generally become shifted,
according to the rule (\ref{sl2r}). In order to restore the moduli to
the original vacuum $\tau_0=i$, while keeping $Q_2=(p_2,q_2)^{\rm T}$
fixed, we must act with a compensating Borel-trombone
transformation.  Indeed such a (modulus and charge dependent) transformation
exists, and is unique.  The detailed form 
of the compensating transformation
needed to return $\tau_0$ to its initial value is somewhat
complicated, and we shall not present its general form here. As an
example, however, one may consider a simpler special case, where the
initial charge vector is $Q_1=(1,0)^{\rm T}$, and the $SL(2,\Z)$
matrix $\Lambda$ mapping from $Q_1$ to $Q_2$ has $a=p_2$, $c=q_2$. For
this special case, the Borel-trombone compensator is
\be 
{\bf B}\, t=\pmatrix{ d p_2+ q_2^2 & -p_2 b -p_2 q_2\cr d q_2 - p_2 q_2&
p_2^2 -b q_2}\ .\label{comp} 
\ee

     In the general case of an $SL(2,\Z)$ mapping between arbitrary $Q_1$ and 
$Q_2$ on the elementary charge lattice, we note that the Iwasawa decomposition 
for $SL(2,\R)$ allows one to factorize $\Lambda\in SL(2,\Z)$ as
\be
\Lambda=\tilde{\bf B}\, {\bf H}\ ,\label{BH}
\ee
where $\tilde{\bf B}$ is an element of the Borel group that leaves $Q_2$
invariant up to scaling and ${\bf H}$ is an element of the stability
group $H=SO(2)$ of the point $\tau_0=i$. This stability group is at
the same time the linearly-realized subgroup of the standard classical
$G=SL(2,\R)$ symmetry group, with the scalar fields taking their
values in $G/H$. Clearly, it is only the $\bf B$ transformation that
actually causes $\tau_0$ to move, so the Borel-transformation part of
the compensator must be simply ${\bf B}= \tilde{\bf
B}^{-1}$. Consequently, the Borel-transformation parts of the
compensator and of $\Lambda$ cancel out, and one is left simply with
\be
{\bf B}\, t\, \Lambda = t\, {\bf H}\ ,\label{tH}
\ee
{\it i.e.}\ the compensated $SL(2,\Z)$ transformation may be realised as a
specific $SO(2)$ transformation $\bf H$ times a trombone rescaling $t$. It 
should be noted here that the matrix ${\bf H}$ (and also the product 
$t\, {\bf H}$) is not generally an integer-valued matrix.

     We shall call the compensated $SL(2,\Z)$ transformations the {\it
active} $SL(2,\Z)$, in order to distinguish them from the standard
supergravity $SL(2,\R)$ transformations which move both the modulus
$\tau_0$ and the charges at the same time, and which in string theory
are discretised to $SL(2,\Z)$ and interpreted as a local, or 
{\it gauged} symmetry. In other words,
states related by the gauged $SL(2,\Z)$ are identified in string
theory.  In fact, the gauged symmetry is what is customarily called the
U-duality symmetry \cite{ht}.  The gauged $SL(2,\Z)$ acts linearly on
the charges and field strengths of the theory, according to the standard
transformation rules. The active $SL(2,\Z)$ transformations, on the
other hand, maintain a fixed vacuum value of the modulus $\tau_0$, and
map the charge vectors between physically-distinct values on the charge
lattice in the given fixed vacuum. The active $SL(2,\Z)$ manages to
avoid moving the scalar modulus by means of the compensation
construction, using specific details of the modulus $\tau_0$ that is to
be maintained and of the final charge $Q$ that is reached (since the
Borel group used is the Borel group for this specific $Q$).
Consequently, the active $SL(2,\Z)$ is in general realised {\it
nonlinearly} on the variables of the theory, not only on the scalar
fields but also on the 3-form field strengths and the metric. When the
action of the active $SL(2,\Z)$ transformations is considered
specifically on the charges $Q$, however, this generally nonlinear
transformation simplifies to the linear transformation (\ref{LQ}). The
action on fields in general, however, is nonlinear. Note that although
the moduli, {\it i.e.}\ the asymptotic values of the scalar fields, are
held fixed, the scalar fields throughout spacetime do generally
transform.  A crucial point, which emphasises the distinction between
the gauged and the active $SL(2,\Z)$ symmetries, is that while the gauged
$SL(2,\Z)$ preserves the value of the mass, the active $SL(2,\Z)$
changes it, since the metric is also transformed.  Although an
identification of states under the gauged symmetry is perfectly
consistent, it would clearly be inconsistent to identify states which
can have different masses under the active symmetry.  The active
$SL(2,\Z)$ is a genuine spectrum-generating symmetry, and should not be
confused with the standard U-duality, which cannot generate the spectrum
since it cannot change the masses.

     The detailed expression for the active transformation can be
given without restricting it to the Dirac-quantized $SL(2,\Z)$, since
the compensation construction given above can be carried out also for
all classically-allowed $SL(2,\R)$ transformations. The key to
deriving the specific form of the $t$ and $\bf H$ parts of this
transformation in (\ref{tH}) is to note that, when acting specifically
on the charges $Q$, the transformation becomes linear, so that in
mapping from a charge vector $Q_i=(p_i,q_i)$ to a charge vector $Q_j$,
corresponding to the application of an $SL(2,\R)$ matrix $\Lambda$,
one must have
\be
t_{ji}\, {\bf H}_{ji}\, Q_i = Q_j = \Lambda Q_i\ ,\label{tHLQ}
\ee
where the $i$, $j$ subscripts are not indices, but labels corresponding to
the charge configurations $Q_i$, $i=1,2,\ldots$ reached. 

     The trombone scaling part of the transformation (\ref{tHLQ}) is
given by
\be
t_{ij}={m_i\over m_j}\ ,\label{tij}
\ee
where $m_i$ is the mass, which is given, in the $\tau_0=i$ vacuum that we
are initially considering here, by
\be
m_i=\sqrt{p_i^2+q_i^2}\ .\label{mass}
\ee
This expression for $m_i$ is an example of an $H$-invariant ``length''
for a charge vector, as referred to earlier, specialised to the
$\tau_0=i$ vacuum. We shall give the corresponding expression for
general $\tau_0$ shortly.

     The $H=SO(2)$ part of the transformation is given by
\be
{\bf H}_{ij}=\pmatrix{\cos\theta_{ij}&\sin\theta_{ij}\cr
-\sin\theta_{ij}&\cos\theta_{ij}}\ ,\qquad 
\theta_{ij}=\theta_i-\theta_j \ ,\label{Hmatrix}
\ee
where $\tan\theta_i = p_i/q_i$.  Note that the product $t_{ij}\, {\bf H}_{ij}$
is given by 
\be
t_{ij}\, {\bf H}_{ij} =\pmatrix{\a &\beta\cr -\beta &\a} \ ,\label{f247}
\ee
where
\be
\a= \fft{p_1\, p_2 + q_1\, q_2}{p_1^2 +q_1^2}\ ,\qquad
\beta = \fft{p_2\, q_1 -p_1\, q_2}{p_1^2 + q_1^2}\ .\label{f248}
\ee

     It is straightforward to generalise the above discussion to the
case where a generic point $\tau_0=\chi_0 + i/g$, rather than
$\tau_0=i$, is taken to be the vacuum modulus ($g=e^{\phi_0}$ is the
string coupling constant).  This can be done by noting that the
stability group $H$ of $\tau_0$ leaves the matrix ${\cal M}_0$
invariant, {\it i.e.}\ ${\cal M}_0 = {\bf H}\, {\cal M}_0\, {\bf
H}^T$, where ${\cal M}_0$ is the matrix ${\cal M}$ defined in
(\ref{mato}), but with the scalar fields replaced by their asymptotic
values $\phi_0 =\log g$ and $\chi_0$, and ${\bf H}$ denotes an element of 
$H$.  We can write ${\cal M}_0 = V\,
V^T$, where the ``vielbein'' $V$ is given by the $SL(2,\R)$ matrix
\be
V=\fft1{\sqrt g} \, \pmatrix{1 & g\, \chi_0 \cr
                             0 & g}\ .\label{viel}
\ee
The previous modulus point $\tau_0=i$ corresponds to ${\cal M}_0 = 1$.
The matrix $V$ (\ref{viel}) may also be viewed as an $SL(2,\R)$
element that maps the vacuum modulus from $i$ to $\tau_0$.  Thus the
stability group matrices for a generic $\tau_0$ will be of the form
(\ref{Hmatrix}), but conjugated with $V$, so that the group element
${\bf H}_{ij}$ is now given by
\be
{\bf H}_{ij} = V\, \pmatrix{\cos\tilde\theta_{ij}&\sin\tilde\theta_{ij}\cr
-\sin\tilde\theta_{ij}&\cos\tilde\theta_{ij}} \, V^{-1}\ ,\qquad 
\tilde\theta_{ij}=\tilde\theta_i-\tilde\theta_j \ ,\label{Hmatrix1}
\ee
where 
\bea
\tilde\theta_{ij}&=&\tilde\theta_i-\tilde\theta_j\label{thetadif}\\
\tan\tilde\theta_i &=& \tilde p_i/\tilde q_i\ .\label{tantheta}
\eea 
The quantities $\tilde Q_i=(\tilde p_i, \tilde q_i)^T$ are related to
the charges\footnote{We continue to take the electric charges
$Q=(p,q)$ to be integers in the generic vacuum.  This is purely for
convenience, and any other choice allowed by the Dirac quantisation
condition will lead to identical conclusions about the nonlinearly-realised
spectrum-generating group. In fact the Dirac
quantisation condition does not by itself completely determine the
lattice of allowed charge states; any $SL(2,\R)$ transformation
of the integer lattice would be equally valid (provided of course that the
lattice of magnetic 5-brane charges is transformed contragediently at
the same time). In order to fix the lattice, one
needs also to specify a set of standard charge states in each given
vacuum. In the present purely electric-charge system, it is natural to
make this choice without regard to the values of the scalar moduli,
following Ref.\ \cite{schwarz}. This choice preserves the symmetry
between the two 3-form field strengths $F^{(i)}_3$, and has the
consequence that the $SL(2,\Z)$ group that preserves the charge
lattice is always represented by integer-valued matrices, irrespective
of the value of the scalar modulus $\tau_0$.}  $Q_i=(p_i,q_i)^T$ by
\be
\tilde Q_i = V^{-1}\, Q_i\ .\label{ch}
\ee  
This implies that
\be
\tan\tilde\theta_i = g\, \tan\theta_i -g\, \chi_0\ .
\ee
The trombone scaling $t_{ij}$ is again given by (\ref{tij}), but the mass
$m_i$ is now given by
\be
m_i^2 = Q_i^T\, {\cal M}_0^{-1}\, Q_i = \tilde Q_i^T\,\tilde Q_i\ .
\label{genmass}
\ee
The product $t_{ji}\, {\bf H}_{ji}$ transforms the charges according to 
(\ref{tHLQ}), while leaving the generic vacuum $\tau_0$ invariant.
       
     The mass defined in (\ref{genmass}) is also the 
$H$-invariant ``length'' of the charge vector for the vacuum specified
by $\tau_0$. It has the property of being preserved in form under $H$
transformations, which hold the moduli fixed. 

\section{Group structure of the active $SL(2,\Z)$}

     As is clear from the fact that the compensated active
transformation constructed above is generated solely by the $SO(2)$
and trombone combination (\ref{tH},\ref{tij},\ref{Hmatrix1}), this
cannot be a linear realisation of $SL(2,\R)$ or
$SL(2,\Z)$. Nonetheless, we have in the continuous-parameter case a
perfectly proper realisation of $SL(2,\R)$, and this may then be
restricted to its discrete $SL(2,\Z)$ subgroup. We shall focus on the
continuous-parameter case in demonstrating that this is an acceptable
group realisation.  The nonlinear dependence of the transformation on
the scalar moduli and on the initial values of the charges
$Q_1=(p_1,q_1)$, as can be seen in (\ref{Hmatrix1}), requires care in
establishing that this is a proper realisation of $SL(2,\R)$.

     In order to have a proper group action of $SL(2,\R)$ on a set
$\sf X$, in which an invertible operator ${\cal O}(\Lambda)$ maps the
point $x\in{\sf X}$ into ${\cal O}(\Lambda)(x)$, it is necessary that
the composition of two such transformations respect the $SL(2,\R)$
group composition rules, {\it i.e.}\
\be
{\cal O}(\Lambda_2)\Big({\cal O}(\Lambda_1)(x)\Big)=
{\cal O}(\Lambda_2\Lambda_1)(x)\ ,
\label{comprel}
\ee
where the product $\Lambda_2\Lambda_1$ is the ordinary matrix product
of two $SL(2,\R)$ matrices. The set $\sf X$ being acted upon here
consists of the full set of fields of type IIB supergravity, since all
of them, including the metric, transform either under the $SO(2)$ part
(\ref{Hmatrix1}) or the trombone part (\ref{tij}) of the
transformation (\ref{tH}).  More explicitly, in a notation that 
indicates the action on the charges, we may write the composition 
law as
\be
{\cal O}(\Lambda_2,\Lambda_1 Q_1)\, {\cal O}(\Lambda_1,Q_1)=
{\cal O}(\Lambda_2 \Lambda_1,Q_1)\ .\label{ccomp}
\ee

     For the nonlinear realisation (\ref{tH}), establishing
the composition property (\ref{comprel}) amounts to verifying that the
trombone part (\ref{tij}) and the $SO(2)$ part (\ref{Hmatrix1})
separately respect the $SL(2,\R)$ group composition rules. We consider
mapping from a configuration labelled 1 to a configuration labelled 2
and then on to a configuration labelled 3. For the trombone part of the
transformation, the check of $SL(2,\R)$ composition is
straightforward:
\be
t_{32}\, t_{21}={m_3\over m_2}\cdot{m_2\over m_1}=t_{31}\ ,\label{tcomp}
\ee
where the masses $m_i$ for general $\tau_0$ are given by (\ref{genmass}). 
For the $SO(2)$ part of the transformation, the check of $SL(2,\R)$ 
composition starts by combining the matrices (\ref{Hmatrix1}):
\bea
{\bf H}_{32}\, {\bf H}_{21}&=&V\, \pmatrix{\cos\tilde\theta_{32}&\sin
\tilde\theta_{32}\cr
-\sin\tilde\theta_{32}&\cos\tilde\theta_{32}} \, V^{-1}\,
V\, \pmatrix{\cos\tilde\theta_{21}&\sin\tilde\theta_{21}\cr
-\sin\tilde\theta_{21}&\cos\tilde\theta_{21}} \, V^{-1}\nonumber\\
&&\nn\\
&=&V\, \pmatrix{\cos(\tilde\theta_{32}+\tilde\theta_{21})&
\sin(\tilde\theta_{32}+\tilde\theta_{21})\cr
-\sin(\tilde\theta_{32}+\tilde\theta_{21})&\cos(\tilde\theta_{32}+
\tilde\theta_{21})} \, V^{-1}\label{Hcomp1}
\eea
Next, the sum of the $SO(2)$ angles $\tilde\theta_{32}+\tilde\theta_{21}$ 
combines to produce a single angle $\tilde\theta_{31}$, as one can see from
(\ref{thetadif}). Thus we have the desired composition law
\be
{\bf H}_{32}\, {\bf H}_{21} = {\bf H}_{31}\ .\label{Hcomp}
\ee
In terms of the notation in the group composition rule (\ref{ccomp}), this
becomes
\be
\Big(\tilde\theta(\Lambda_2\Lambda_1 Q_1) - 
\tilde\theta(\Lambda_1 Q_1)\Big) +
\Big(\tilde \theta (\Lambda_1 Q_1) - \tilde \theta(Q_1)\Big)
=\tilde\theta (\Lambda_2\Lambda_1 Q_1) -
\tilde \theta(Q_1)\ ,
\ee
where the $\td \theta(Q)$ is given by (\ref{tantheta}) and (\ref{ch}).

     Consequently, when acting on any field of the type IIB
supergravity theory, the non\-linearly-realised transformation
(\ref{tH}) respects the $SL(2,\R)$ group composition rule, so we do in
fact have a proper $SL(2,\R)$ realisation. One may then simply
restrict this to the $SL(2,\Z)$ discrete subgroup in order to obtain
the transformations that map between the allowed solutions
corresponding to points on the Dirac-quantised elementary charge
lattice.  

    It should again be emphasised that the discretisation of the
active spectrum-generating $SL(2,\R)$ is a consequence purely of the
Dirac quantisation condition, and involves no additional input. The
discretisation of the standard $SL(2,\R)$ symmetry to the conjectured
$SL(2,\Z)$ U-duality group is quite a different matter, however, and
is one that can be settled only with some additional input, for
example from string theory; the Dirac quantisation condition for BPS
states is not by itself enough to imply a discretisation of the
standard $SL(2,\R)$.  The fact that in the type IIB theory the
standard $SL(2,\R)$ is a purely non-perturbative symmetry makes the
issue of its discretisation particularly tricky to study, and
emphasises the importance of not confusing it with the active
spectrum-generating $SL(2,\R)$, whose discretisation is easily
established.  In order to clarify this point, let us try to see at
what stage one is in a position to deduce that the standard $SL(2,\R)$
symmetry is discretised to $SL(2,\Z)$.  As we have previously
observed, the Dirac quantisation condition itself can be satisfied by
any charge lattice that is generated by an arbitrary $SL(2,\R)$
transformation of a given allowed lattice.  On the other hand, at the
classical level we have a global $SL(2,\R)$ symmetry that allows us to
view a modulus point $\tau_0^i$ with charge lattice $C^i$ as merely a
relabelling of a modulus point $\tau_0^j$ with charge lattice $C^j$ if
the moduli and charges are simply related to one another by an
$SL(2,\R)$ symmetry transformation. Thus to avoid an overcounting of
$(\tau_0^i,C^i)$ pairs, we should mod out by this $SL(2,\R)$, viewed
as a relabelling symmetry, and only count as distinct those pairs that
are not related by $SL(2,\R)$ symmetry transformations.  In other
words, the set of all possible charge lattices above every point on
the $SL(2,\R)/SO(2)$ modulus space is equivalent to a principle fibre
bundle with an $SL(2,\R)$ fibre above each modulus point. Modding out
by the relabelling amounts to choosing a cross-section of this bundle.

     Different choices of cross-section should not generally be
thought of as being physically equivalent. In particular, different
choices, all equally allowed by the Dirac quantisation condition, can
give differing unbroken residual symmetries. For example, there is a
family of cross-sections obtained by taking an integer charge lattice
at some given modulus point, and then, at other modulus points that
are obtained from the given one by operation with an $SL(2,\R)$
transformation $\Lambda$, choosing the charge lattice to be $\Lambda$
rotations of the initial integer lattice. Such a cross-section leaves
the full classical $SL(2,\R)$ symmetry unbroken, by
construction. Another example is to choose the {\it same} charge
lattice $C$ for every point $\tau_0^i$ in the modulus space.  In this
case, there is a residual $SL(2,\Z)$ symmetry of the $(\tau_0^i,C)$
pairs. The specific charge lattice chosen in this second example need
not be an integer lattice, but in this case, there will be a
relabelling of the $(\tau_0^i,C)$ pairs {\it via} a fixed $SL(2,\R)$
transformation that transforms the charge lattices into integer
lattices.  The surviving $SL(2,\Z)$ symmetry for a given cross-section
of this type will be given by conjugation of the integer-valued
representation of $SL(2,\Z)$ by the fixed $SL(2,\R)$ element. In such
a case, with a surviving $SL(2,\Z)$, one has the option of dividing
out by this surviving discrete symmetry, an option that has been
argued to be taken up in string theory. This has the effect of
restricting the modulus space to its fundamental domain. 

     For a generic cross-section, the classical $SL(2,\R)$ will be
completely broken down to the identity. The conclusion is that the
standard classical $SL(2,\R)$ symmetry is not discretised by virtue of
the Dirac quantisation condition alone. In order to determine the
unbroken symmetry, one must choose a cross-section of the bundle of
lattices over modulus space. This choice must be made using additional
information over and beyond the Dirac quantisation
condition. Frequently in the litterature, one encounters the
additional requirement that purely electric states exist in the
lattice, for example. And string theory introduces modifications to
the classical theory that appear to select the $SL(2,\Z)$-preserving
cross-sections. Evidence for the discretisation of the standard
$SL(2,\R)$ in string theory has been offered in \cite{gvh}, where
candidate counterterms that preserve only an $SL(2,\Z)$ subgroup of
the standard $SL(2,\R)$ have been given.

     In referring to the $SL(2,\Z)$ ``gauged'' subgroup of the
standard supergravity $SL(2,\R)$ in this paper, we are making an
implied choice of the second kind of lattice-bundle cross-section
described above, for which the same lattice of canonical charges is
chosen at each point in modulus space. Then, by an $SL(2,\R)$
transformation and relabelling of the points of modulus space, these
lattices may also be taken to be integer-valued. This choice appears to
be the standard one in string-theory discussions \cite{schwarz}, but 
the additional assumptions lying beneath this choice should be more
carefully inspected.

\section{Group-theoretical structure of the spectrum}

     The active $SL(2,\R)$ transformation acts transitively on the
charge-vector space of the type IIB theory's string soliton solutions
in a given vacuum. This space is the two-dimensional Euclidean plane
with the origin excluded, $\E^2\backslash\{\vec0\}$ (the origin is
excluded if one wants to focus attention only on solutions having the
same degree of unbroken supersymmetry, thus excluding the zero-charge
pure Minkowski-space solution). This transitive group action makes it
straightforward to identify the spectrum from a group-theoretical
perspective. Whenever a group's orbit coincides with the whole of the
set on which it acts, {\it i.e.}\ whenever the group acts transitively
on the set, the realisation is equivalent to that on a coset space
$G/S_{x_0}$, where $S_{x_0}$ is the stability group of any particular
chosen point $x_0$ on the orbit.

     In the classical $SL(2,\R)$ case with a continuous charge-vector
space, the stability group of a given charge $Q$ is isomorphic to the
strict Borel group, $S_Q\simequiv B_{\rm strict}$, whose entries are
purely upper-triangular ({\it i.e.}\ excluding the Cartan
subalgebra). For the present case with $G=SL(2,\R)$, the coset space
$SL(2,\R)/B_{\rm strict}$ is indeed equivalent to
$\E^2\backslash\{\vec0\}$. In order to see this, consider the
equivalence relation implied by membership in a given $SL(2,\R)/B_{\rm
strict}$ coset. Taking the charge vector $(1,0)$ to correspond to the
chosen point $x_0$ above, the corresponding stability group has
elements
\be 
B_{\rm strict}=\pmatrix{1&k\cr0&1}\ ,\label{Bstr} 
\ee
where $k\in\R$ is an arbitrary real number.  The equivalence relation
between coset members is then 
\bea 
\pmatrix{a&b\cr
c&d}&\sim&\pmatrix{a&b\cr c&d}\,\pmatrix{1&k\cr 0&1}\nn\\ &&\nn\\
&=&\pmatrix{a&ak+ b\cr c&ck+d}\ ,\label{sl2rcoset} 
\eea 
implying that {\it all} $SL(2,\R)$ matrices with given $a$ and $c$ are
equivalent.  The natural label for this equivalence class is the
vector $(a,c)$, whose entries cannot both vanish because $a$, $b$,
$c$, $d$ must satisfy $ad-bc=1$. (In the discrete case, the
equivalence relation holds for all $SL(2,\Z)$ matrices with given $a$
and $c$.) Indeed, this labelling establishes an equivariant map
between $SL(2,\R)/B_{\rm strict}$ and $\E^2\backslash\{\vec0\}$, thus
establishing the equivalence.  (In other words, this is a one-to-one
map between $SL(2,\R)/B_{\rm strict}$ and $\E^2\backslash\{\vec0\}$
that preserves the action of the $SL(2,\R)$ group.)

     In the quantum case, where the active symmetry group is reduced
to $SL(2,\Z)$, the space of states allowed by the Dirac quantisation
condition is the charge lattice $(p,q)$ with $p$ and $q$ non-vanishing
integers. The action of $SL(2,\Z)$ on this set is not transitive,
however. As we have observed earlier, the various disjoint orbits of
this discrete group may be characterised by the points $(n,0)$ that
they contain. We have identified the irreducible orbit containing the
point $(1,0)$ as the ``elementary'' orbit. When the charge lattice is
restricted to this sub-lattice, for which the integers $p$ and $q$ are
always relatively prime, the action of $SL(2,\Z)$ once again becomes
transitive. We may then make a group-theoretical identification of
this elementary discrete orbit. The standard charge vector $(1,0)$
lies on the elementary $SL(2,\Z)$ orbit, and for this chosen point the
stability group $B_{\rm strict}(\Z)$ is once again of the form
(\ref{Bstr}), but with $k$ now restricted to be an integer,
$k\in\Z$. Then the equivalence relation between coset elements in
$SL(2,\Z)/B_{\rm strict}(\Z)$ is once again of the form
(\ref{sl2rcoset}), but now with $k\in\Z$. The natural label for this
equivalence class is again $(a,c)$, but with $a$ and $c$ now
integers. These charge vectors coincide precisely with the points on
the elementary $SL(2,\Z)$ orbit because $SL(2,\Z)$ matrices cannot
satisfy the constraint $ad-bc=1$ unless the integers $a$ and $c$ are
relatively prime.

     In the continuous-charge classical case, the strict Borel group
(\ref{Bstr}) is isomorphic to $\R$, so the classical charge spectrum
may be identified as $SL(2,\R)/\R$. In the quantum case, the strict
Borel group is isomorphic to $\Z$, so the elementary orbit of string
solitons may be identified as $SL(2,\Z)/\Z$.

\section{Lower-dimensional cases}

     Up until now, our discussion has focussed on the type IIB supergravity
in $D=10$.  As we shall now show, many of the features that we encountered
there persist in lower-dimensional examples.  However, there are also some
additional subtleties that need to be considered.

     In the $D=10$ type IIB case, the only transformation beyond the
standard $SL(2,\R)$ symmetry that was needed in order to construct the
active symmetry group that leaves the scalar modulus fixed was a single
trombone scaling transformation. Indeed, this trombone symmetry of the
equations of motion is the only readily available symmetry that one
has in any dimension which can be used in the compensation construction
of such fixed-modulus active transformations.  Although it might not
seem immediately apparent that this is all that is necessary in lower
dimensions, where the symmetry groups are larger, this does in fact
prove to be the case, at least for the construction of symmetries
acting on multiplets of ``fundamental'' supergravity solitons, in a
sense that we shall now define.

     The key point to recognise in dealing with the lower-dimensional
cases is that for any of the maximally-noncompact supergravity
symmetry groups $G$ shown in Table \ref{tab:sugrasyms}, one has an Iwasawa
decomposition of a general group element $\Lambda$, specialised to the
vacuum point on the scalar modulus manifold $G/H$ and to the charges
$Q$ defining a given fundamental soliton solution:
\be
\Lambda={\bf B}\,{\bf H}\ ,\label{genLBH}
\ee
where $\bf H$ is an element of the stability group $H$ of the vacuum
modulus point ${\cal M}_0$ on $G/H$ and $\bf B$ is an element of the
Borel subgroup corresponding to $Q$. (The Iwasawa decomposition and
the Borel subgroups (which leave highest weight vectors invariant up
to rescaling) of the global supergravity symmetry groups
$E_{11-D}$ were extensively studied in \cite{sol}.)  In writing this
Iwasawa decomposition, we are making an important assumption that the
supergravity $p$-brane soliton is ``fundamental,'' in the sense of
lying on the same symmetry orbit as a single-charge solution, for
which only one of the theory's field strengths is non-vanishing. In
this case, $\bf B$ can be taken to belong to a subgroup of $G$ that is
isomorphic to the canonical upper-triangular Borel subgroup, {\it via}
a similarity transformation using an appropriate element of $H$, such
that all the elements of this subgroup leave the charge $Q$ invariant
up to an overall scaling. In this case, it is clear that all that is
necessary to construct a nonlinearly-realised active symmetry
transformation is a single trombone transformation that can compensate
for the overall scaling of $Q$.  Indeed, even though we have more
complicated group structures in lower dimensions, it should be noted
that there exists precisely one independent trombone symmetry, since
if there were two, we could find one combination that left the
Einstein-frame metric invariant, and hence would be part of the
standard supergravity global symmetry group.

\begin{table}[ht]
\centering
\vspace{0.4cm}
\begin{tabular}{|c|c|c|}
\hline
$D$&$G$&$H$\\
\hline\hline
9&$GL(2,\R)$&$SO(2)$\\
\hline
8&$SL(3,\R)\times SL(2,\R)$&$SO(3)\times
SO(2)$\\
\hline
7&$SL(5,\R)$&$SO(5)$\\
\hline
6&$SO(5,5)$&$SO(5)\times SO(5)$\\
\hline
5&$E_{6(+6)}$&$U\!Sp(8)$\\
\hline
4&$E_{7(+7)}$&$SU(8)$\\
\hline
3&$E_{8(+8)}$&$SO(16)$\\
\hline
\end{tabular}
\caption{Supergravity symmetry groups.\label{tab:sugrasyms}}
\end{table}

     The restriction to fundamental solutions in this discussion
derives from the requirement that there exist a Borel subgroup leaving
the charge configuration $Q$ invariant up to an overall scaling. If
only a single charge component is turned on, it is easy to see that
such a Borel subgroup will exist. Moreover, such a Borel subgroup will
exist for any charge configuration on the same orbit as the
single-charge configuration, obtainable by an appropriate similarity
transformation. To see this in detail, we first note that the
single-charge solutions supported by field strengths of a given rank
form an irreducible representation under the Weyl group \cite{weyl} of
the standard supergravity symmetry group as shown in Table
\ref{tab:sugrasyms}. One of these single-charge solutions will be a
highest-weight state \cite{cjlp}, say with charge $Q_{\rm h}$, and
will therefore be invariant up to scaling under the canonical
upper-triangular Borel group\footnote{The algebra of the canonical
Borel group comprises the positive-root generators and the Cartan
generators of the supergravity symmetry group.} $B_{\rm c}$. Any
solution, with charge $Q=\Lambda Q_{\rm h}$, lying on the same orbit
as the highest-weight solution, will have a corresponding Borel group
$B$, obtained from the canonical one by the similarity
transformation\footnote{Note that if one makes an Iwasawa
decomposition of $\Lambda$ into an element of the vacuum stability
group $H$ and an element of $B_{\rm c}$, then only the element of $H$
is effective in moving $B_{\rm c}$ into $B$, so the similarity
transformation rotating $B_{\rm c}$ into $B$ is effectively made by an
element of $H$.} $B=\Lambda B_{\rm c}\Lambda^{-1}$.

     Once one has made the decomposition (\ref{genLBH}) of a general
group element, the Borel-group factor is the one that should be
cancelled out in constructing the compensated active symmetry
transformation. In order to see that the Borel group $B$ is precisely
the part of $G$ that is effective in moving the scalar moduli, note
that the scalar manifold $G/H$ can be parametrised in the fashion
\be
{\cal M}=VV^\#\ ,
\ee
where $V^\#=\tau(V^{-1})$ and $\tau$ is the Cartan involution, whose
fixed-point set is the maximal compact subgroup $H$. In simple cases,
where $H$ has a regular embedding in $G$, 
$V^\#=V^{\rm T}$ if $H$ is orthogonal, $V^\#=V^\dagger$ if $H$ is
unitary and $V^\#=\Omega V^\dagger$ if $H$ is a $U\!Sp$ group, with
$\Omega$ its invariant symplectic matrix. The matrix $V$ generalises
the vielbein (\ref{viel}) of the $SL(2,\R)$ case, and is an element of
the Borel group.  The scalar moduli are then characterised by the
matrix ${\cal M}_0$, which is the asymptotic form of ${\cal M}$.
Thus, the Borel group is precisely what is needed in order to move the
scalar moduli around on $G/H$, since given an element ${\bf B}\in B$,
the matrix $\cal M$ (which in general transforms according to ${\cal
M}\rightarrow \Lambda{\cal M}\Lambda^\#$) transforms to
\bea
{\cal M}'&=&{\bf B}{\cal M}{\bf B}^\#\nn\\
         &=&{\bf B}V({\bf B}V)^\#\ ,
\eea
so that $V$ transforms to ${\bf B}V$, which is just another element of the
same Borel group.

     In the solitonic-string example that we began with in $D=10$ type
IIB supergravity, all of the solitonic solutions are of the
fundamental type. In dimensions 9 and lower, however, ``multi-charge''
solutions are also found, for which the construction of an active
symmetry transformation using a single trombone scaling transformation
is not possible. These multi-charge solutions\footnote{In the
classification of Refs \cite{stainless,multicharge}, the single-charge
solutions are characterised by a parameter $\Delta$ that takes the
value 4, and multi-charge
solutions are characterised by $\Delta=4/N$, with $N>1$.} are
all characterised by a lower degree of supersymmetry preservation than
the fundamental solutions. The fundamental solutions all preserve 1/2
of the rigid supersymmetry of the supergravity theory, but the
multi-charge solutions preserve 1/4 or less. Such multi-charge
solutions also possess static generalisations in which the charge centers
corresponding to the independent field strengths are separated, so the
multi-charge solutions may be interpreted as ``bound states at
threshold'' of single-charge solutions \cite{bs1,bs2,bs3}. It is not
yet known whether there is a group-theoretical origin to the various
moduli of the multi-charge solutions. If there is, including a
transformation that generalises the trombone symmetry of this paper
but such that the various independent charge components are
independently scaled, then the above compensation construction of a
fixed-modulus active symmetry could also be carried out for the
multi-charge solutions.

\section{Anomalies}

     The trombone scaling symmetry that forms a crucial ingredient in
the above discussion arose as a symmetry of the classical equations of
motion.  If the BPS soliton multiplets are to survive at the quantum
level, it is important to establish that there is no scaling anomaly
that destroys the trombone symmetry. However, this symmetry is
extremely sensitive to quantum corrections. For generic solutions of
the classical equations, it is clear that quantum counterterms will
certainly spoil this symmetry, because they have different scaling
dimensions from that of the classical Lagrangian. Nonetheless, the
possibility of having soliton multiplets incorporating the trombone
scaling is not by this fact immediately ruled out for the
BPS-saturated $p$-brane solutions that we have been studying.

     The relevant question for the BPS solutions is whether they
persist as solutions with arbitrary scale size in the
quantum-corrected effective equations of motion. To illustrate
the problem, consider the first dangerous perturbative counterterm, {\it
i.e.}\ the first one that cannot be absorbed just by a renormalisation
of the fields of the effective theory. This is the first supersymmetric
invariant that does not vanish when subjected to the classical-level
equations of motion. The counterterm contains a term quartic in
Riemann tensors, plus superpartners for the other fields; the purely
gravitational part of the invariant is the square of the Bel-Robinson
tensor \cite{dks}. In all extended supergravities, there are
extended-supersymmetric generalisations of this counterterm
\cite{dk,kallosh81,hl,hstsa,hstsc}. In all $D=4$ supergravity cases,
these counterterms are expected to occur with infinite coefficients at
the three-loop order \cite{hstmir}; in the sigma-model beta-function
approach to string effective field theories, the corresponding finite
contributions to the effective action occur at the $(\alpha')^3$ order
\cite{gdvz}.

     In $N=2$, $D=4$ superfields, the quartic counterterm takes the form 
\cite{kallosh81,hl}
\be
\Delta I_3 = \int d^4x d^8\theta E W_{\alpha\beta}W^{\alpha\beta}
\bar W_{\dot\alpha \dot\beta}\bar W^{\dot\alpha \dot\beta}\ ,
\label{3loopct}
\ee 
where $E$ is the determinant of the supervielbein,
$W_{\alpha\beta}$ is the $N=2$ supergravity conformal-field-strength
superfield, and all two-component indices $\alpha$, $\dot\alpha$ are
referred to the Lorentz-covariant tangent space. As one can see, this
$D=4$ counterterm scales like $\lambda^{-2}$ under the trombone
scaling (\ref{d11trombone}), whereas the $D=4$ classical action scales
as $\lambda^2$, so the presence of this counterterm certainly ruins
the trombone symmetry as far as general field configurations are
concerned. Nonetheless, one might still hope to find the integration
constants, implied at the classical level by the trombone symmetry
(\ref{d11trombone}), to be present in {\it certain classes} of
solutions to the full theory, such as the class of BPS states.

     The question of whether the counterterm (\ref{3loopct}) vanishes
in BPS backgrounds has been discussed in \cite{kallosh92}, where it
was shown that the counterterm itself vanishes for one of the classic
BPS solutions, {\it i.e.}\ for the original extreme Reissner-Nordstrom
solution \cite{hawkel}. Unfortunately, it is not enough to establish
just that counterterms themselves vanish in BPS backgrounds, so the
analysis of Ref. \cite{kallosh92} remains incomplete.  In order for
the BPS solutions to be stable against quantum corrections, it is
necessary to ensure that the {\it variations} of counterterms vanish
in BPS backgrounds, {\it i.e.}\ that the BPS field configurations
remain good solutions when substituted into the counterterm-corrected
equations of motion.

     We shall not carry out a full nonlinear analysis of this question
for the counterterm (\ref{3loopct}), but shall be content to indicate
how its variation manages to vanish for BPS backgrounds using a
linearised superfield analysis \cite{hstsc} considering, for example, an
asymptotic region where the deviations from flat empty space are small.
This analysis is actually made simpler by the dangerousness of the
counterterm (\ref{3loopct}), which is built using precisely those
field-strength combinations that do not vanish when subjected to the
classical Einstein-supergravity field equations, {\it i.e.}\ from the
field strengths of $D=4$, $N=2$ {\it conformal} supergravity. Of course,
as one can see from its non-trivial scaling dimension, (\ref{3loopct}) is
not really a conformal supergravity invariant. However, at the leading
order in deviations from flat space, it becomes a linearised
superconformal invariant, and so it depends to this order only on the
non-gauge parts of the $N=2$ supergravity multiplet with respect to
$N=2$ conformal supergravity. At the linearised level, the superfield
$W_{\alpha\beta}$ may be expressed in terms of an $N=2$ conformal
supergravity prepotential $V$ \cite{hstsc,gs82}
\be
W_{\alpha\beta}=i\bar D^4D_{\alpha\beta}V\ , \label{linW} 
\ee
where $D_{\alpha\beta}$ is given in terms of the basic superspace
covariant derivatives by $D_{\alpha\beta}=\ft12\epsilon_{ij}
D_\alpha^iD_\beta^j$. In varying the lowest-order term of
(\ref{3loopct}), which is quartic in small quantities and thus depends
only on the linearised superconformally-invariant $W_{\alpha\beta}$
superfield, it is sufficient to vary the conformal prepotential $V$
({\it i.e.}\ the variation of (\ref{3loopct}) at lowest order can have
contributions only from the variables actually present in
(\ref{3loopct}) at that order, and these are the non-gauge parts of
the conformal supergravity multiplet). The result is a contribution to
the equation of motion for the singlet scalar auxiliary field of the
$N=2$ multiplet \cite{dewit80} ({\it i.e.}\ to the field conjugate to
the lowest component of $V$).  Using the supergravity constraints and
equations of motion $\bar D_{\dot\alpha\,i}W_{\alpha\beta}=0$,
$D_\alpha^iW^{\phantom i}_{\beta\gamma} = 
D_{(\alpha}^iW^{\phantom i}_{\beta\gamma)}$ (where the
brackets denote symmetrisation with strength one), for a purely
bosonic field configuration, together with the condition
characterising the BPS extreme Reissner-Nordstrom solution
\cite{kallosh92}
\be
C_{\alpha\beta\gamma\delta}=-\nabla_{(\alpha}^{\dot\rho}
W_{\beta\gamma}^{\phantom{\dot\rho}}
\hat K_{\delta)\dot\rho}^{\phantom{\dot\rho}}\ ,\label{bpsrel} 
\ee 
where $\hat K_{\alpha\dot\beta}$ is the normalised timelike Killing
vector of the static BPS solution with indices referred to the tangent
space, one straightforwardly verifies that the variation of
(\ref{3loopct}) with respect to $V$ vanishes.

     The $D=4$ extreme Reissner-Nordstrom solution thus manages to
escape from this threat of a perturbative anomaly in the trombone scaling
symmetry, but the anomalies arising from (\ref{3loopct}) and from
higher-order terms certainly do affect more general non-BPS solutions.
In fact, the extreme Reissner-Nordstrom black-hole solution is known
from $d=2$ string sigma-model considerations to give a fully
conformally invariant  sigma model \cite{tseytlinreview}, thus managing
to be an exact solution to the full effective supergravity field theory
with arbitrary scale size to all orders in $\alpha'$. This example serves
to illustrate how important the BPS saturation conditions are for the
existence of duality multiplets. In view of the anomalies in the
trombone scaling symmetry at the quantum level for general field
configurations, there is no reason to expect non-BPS solutions to form
duality multiplets such as the $SL(2,\Z)$ lattice that describes the
BPS solutions.

\section{Conclusion}

      In this paper, we have shown that the full set of integration
constants for fundamental BPS supergravity $p$-brane solutions
preserving half of the supersymmetry may be associated to symmetries
of the theory. The essential element that goes beyond the standard
supergravity symmetry groups $G$ shown in Table \ref{tab:sugrasyms} is
the trombone scaling transformation; this allows one to reach
solutions at different mass levels by the application of symmetry
transformations. Provided attention is restricted to the fundamental
solutions preserving half of the supersymmetry, this single extra
transformation is all that is necessary in order for one to be able
reach the full BPS parameter space, for supergravity in any spacetime
dimension. For ``multi-charge solutions'' preserving less than half of
the supersymmetry, it remains unclear whether the full parameter space
of the solutions can be reached by symmetry transformations. However,
since the multi-charge solutions may be interpreted as superpositions
of the fundamental ones, the most essential class of BPS $p$-branes
may be comprehensively discussed from a group-theoretical basis.

      The moduli of the scalar fields of a supergravity theory form a
special class of integration constants, for they include the various
vacuum angles and coupling constants for the theory in a given vacuum,
and in effect define the vacuum. When one restricts attention to a
given vacuum, the BPS solutions are entirely characterised by their
charges. We have shown that there exists a nonlinear realisation of
the symmetry group $G$, which is quite distinct from the standard one
in that it makes essential use of the trombone scaling transformation
and hence can map between solutions at different mass levels, while
transforming the scalar fields in such a way as to hold fixed their
asymptotic values, {\it i.e.}\ while holding the scalar moduli
fixed. This nonlinearly realised symmetry is distinguished from other
ways of covering the charge-vector space in that it allows a
restriction at the quantum level to a discretised group $G(\Z)$ that
maps only between states permitted by the Dirac quantisation
condition.  This nonlinear realisation of $G(\Z)$ is thus the true
spectrum-generating symmetry of the theory, which we have called the
``active'' $G(\Z)$.

      The only known way consistently to quantise supergravity
theories is superstring theory, of course, so one needs to take into
account the string-theory modifications to the above group-theoretical
picture. One important class of such modifications is the infinite
series of corrections to the effective field theory Lagrangian, of the
same general forms as the infinite counterterms of supergravity when
quantised as a field theory, but with individually finite coefficients
when obtained from superstring theory. These perturbative counterterms
have different scaling behaviour from the classical supergravity
Lagrangian, and hence threaten the existence of the active $G(\Z)$
multiplets. When attention is restricted to BPS solutions, however, we
have argued that the active $G(\Z)$ multiplet structure is maintained,
because at least this class of solutions to the effective theory
persists with arbitrary scale size even in the presence of the
perturbative counterterms. What happens in the light of
non-perturbative corrections to the effective field theory is a
different question that should also be carefully considered. We cannot
currently shed much light on this question, but would hope that the
active discrete $G(\Z)$ multiplet structure would persist also in the
face of the non-perturbative corrections.  Of course if the scaling
symmetry were to be broken for BPS states, this would be a problem not
only for the trombone symmetry {\it per se}, but for the whole idea of the
existence of $G(\Z)$ multiplets of BPS states in a given vacuum.

      Another important modification that superstring theory makes to
the above group-theoretical picture is the identification of states
related by the standard $G(\Z)$ transformations which move the
$p$-brane charges and also the scalar moduli at the same time. Because
these standard $G(\Z)$ transformations are interpreted in this local,
or gauged fashion, we have called this group the ``gauged''
$G(\Z)$. Such an identification is possible but not required at the
level of supergravity field theory, but at least those parts of this
group that coincide with string theory T-duality transformations need
to be given a local interpretation, expressing identifications between
modulus/charge configurations related by such transformations. In
Ref.\ \cite{ht}, it was suggested that this local interpretation be
extended to the full gauged $G(\Z)$ group. This local interpretation
marks another important distinction to be drawn between the gauged
$G(\Z)$ and the spectrum-generating active $G(\Z)$ that we have
introduced. After the identifications, the gauged $G(\Z)$ orbits
consist of single points, whereas the active $G(\Z)$ continues to map
transitively around the entire charge lattice of elementary BPS states
in a given vacuum. As for the active $G(\Z)$ symmetry, one must also
consider whether the {\it gauged} $G(\Z)$ symmetry survives quantum
corrections.  At first sight this symmetry, like the trombone
symmetry, might also seem to be at risk from quantum anomalies. If one
wishes to identify states under the gauged $G(\Z)$ then it is
important that it survive as a symmetry of the full theory, and not
merely when restricted to BPS solutions. Evidence that $G(\Z)$ may
survive in non-perturbative string effective field theories, while the
standard continuous classical $G$ symmetry is broken, has recently
been offered in \cite{gvh}.

      The string-theory-induced identifications may have hidden the
necessity for the active $G(\Z)$ transformations that we have
described.  Indeed, the identification of scalar moduli under the
gauged $G(\Z)$ in the true vacuum sector of the theory, {\it i.e.}\
for vanishing $p$-brane charges, or in other words for flat-space
metrics, makes it tempting to simply overlook the requirement of
coordinated transformations of charges at the same time as scalar
moduli that characterises the standard gauged $G(\Z)$
transformations. For example in Ref.\ \cite{dyonic}, a proposal was
made to generate the charge lattice of BPS solitons using just the
gauged $G(\Z)$, by effectively declaring that the scalar moduli being
mapped between are identified, while the charges being mapped at the
same time are not. We must admit to being puzzled by this proposal,
since the only consistent way to implement a discrete group such as
$G(\Z)$ as a gauged or local symmetry is to declare that the states
are actually equivalence classes with respect to the action of the
group. Since this action maps the charges and scalar moduli together,
it would seem that only the definition of equivalence classes under
this joint action could be mathematically consistent. The distinct and
independent active $G(\Z)$ transformations introduced in the present
paper make unnecessary such a construction, in any case. Moreover, the
high degree of sensitivity of the active $G(\Z)$ transformations to
quantum corrections, owing to their trombone-scaling-transformation
content, makes clear a feature that would not otherwise be prominent:
only the BPS solutions enjoy a degree of protection from quantum
anomalies that would otherwise seem certain to obliterate the $G(\Z)$
multiplet structure.

\section*{Acknowledgements}

          We are grateful to Marcus Bremer, Ashok Das, Gary Gibbons,
Chris Isham, Bernard Julia, Igor Lavrinenko, George Papadopoulos and
Paul Townsend for helpful discussions.  H.L. and C.N.P. are grateful
for hospitality at Imperial College, London and SISSA, Trieste.
K.S.S. is grateful for hospitality at ENS, Paris and SISSA, Trieste.


\begin{thebibliography}{99}

\bibitem{cj1} E. Cremmer and B. Julia, {\sl The $N=8$ supergravity
theory-1-the Lagrangian,} Phys. Lett. {\bf B80} (1978) 48; {\sl The
$SO(8)$ supergravity,} Nucl. Phys. {\bf B156} (1979) 141.

\bibitem{cj2} B. Julia, {\sl Group disintegrations;}\\ E. Cremmer, 
{\sl Supergravities in 5 dimensions,} in ``Superspace and Supergravity'',
Eds. S.W. Hawking and M. Rocek (Cambridge Univ. Press, 1981) 331; 267.

\bibitem{ht} C.M. Hull and P.K. Townsend, {\sl Unity of superstring
dualities,} Nucl. Phys. {\bf B294} (1995) 196: hep-th/9410167. 

\bibitem{sen} A. Sen, {\sl Electric and magnetic duality in string
theory,} Nucl. Phys. {\bf B404} (1993) 109: hep-th/9207053;
{\sl  $SL(2,\Z)$ duality and magnetically charged strings,}
Int. J. Mod. Phys. {\bf A8} (1993) 5079: hep-th/9302038.

\bibitem{cjs} E. Cremmer, B. Julia and J. Scherk, {\sl Supergravity
theory in eleven dimensions,} Phys. Lett. {\bf B76} (1978) 409.

\bibitem{cdf} L. Castellani, R. D'Auria and P. Fr\'e, {\sl
Supergravity and superstrings, a geometric perspective. Vols 1-3},
World Scientific, Singapore (1991).

\bibitem{bbo} E. Bergshoeff, C.M. Hull and T. Ortin,
{\sl Duality in the type II superstring effective action}, 
Nucl. Phys. {\bf B451} (1995) 547: hep-th/9504081. 

\bibitem{IIB} M.B. Green and J.H. Schwarz, {\sl Extended supergravity
in ten-dimensions,} Phys. Lett. {\bf B122} (1983) 122;
\newline
J.H. Schwarz and P.C. West, {\sl Symmetries and transformations of
chiral N=2 D=10 supergravity,} Phys. Lett. {\bf B126} (1983)
301;\newline 
J.H. Schwarz, {\sl Covariant field equations of chiral
$N=2$ $D=10$ supergravity,} Nucl. Phys. {\bf B226} (1983) 269;\newline
P.S. Howe and P.C. West, {\sl The complete N=2 D=10 supergravity,}
Nucl. Phys. {\bf B238} (1984) 181.

\bibitem{schwarz} J.H. Schwarz, {\sl An $SL(2,\Z)$ multiplet of type
IIB superstrings,}  Phys. Lett. {\bf B360} (1995) 13;
Erratum-ibid. {\bf B364} (1995) 252: hep-th/9508143.

\bibitem{weyl} H. L\"u, C.N. Pope and K.S. Stelle, {\sl Weyl group
invariance and $p$-brane multiplets,} Nucl. Phys. {\bf B476} (1996)
89: hep-th/9602140.

\bibitem{vertical} H. L\"u, C.N. Pope and K.S. Stelle, {\it Vertical
versus diagonal dimensional reduction for $p$-branes,}
Nucl. Phys. {\bf B481} (1996) 313: hep-th/9605082.

\bibitem{lorentz} P.S. Howe, A.S. Galperin and K.S. Stelle, {\it The 
superparticle and the Lorentz group,} Nucl. Phys. {\bf B368} (1992) 248:
hep-th/9201020.

\bibitem{gvh} M.B. Green and P. Vanhove, {\sl D-instantons, strings and
M-theory}: hep-th/9704145.

\bibitem{cjlp} E. Cremmer, B. Julia, H. L\"u and C.N. Pope, work in
progress.

\bibitem{stainless} H. L\"u, C.N. Pope, E. Sezgin and K.S. Stelle, 
{\sl Stainless super $p$-branes,} Nucl. Phys. {\bf B456} (1995) 669:
hep-th/9508042.

\bibitem{multicharge} H. L\"u and C.N. Pope, {\sl $p$-brane solitons
in maximal supergravities,} Nucl. Phys. {\bf B465} (1996) 127:
hep-th/9512012; {\sl Multi-scalar $p$-brane solitons,}
Int. J. Mod. Phys. {\bf A12} (1997) 437: hep-th/9512153.

\bibitem{sol}  L. Andrianopoli, R. D'Auria, S. Ferrara, P. Fr\'e, 
and M. Trigiante, {\sl R-R scalars, U-duality and solvable Lie algebras},
Nucl. Phys. {\bf B496} (1997) 617: hep-th/9611014;\\
L. Andrianopoli, R. D'Auria, S. Ferrara, P. Fr\'e, R. Minasian and 
M. Trigiante, {\sl Solvable Lie algebras in type IIA, type IIB and M
theories}, Nucl. Phys. {\bf B493} (1997) 249: hep-th/9612202;\\
L. Andrianopoli, R. D'Auria, S. Ferrara, P. Fr\'e and M. Trigiante,
{\sl $E_{7(7)}$ Duality, BPS black hole evolution and fixed
scalars}: hep-th/9707087;\\
S. Ferrara and J. Maldacena, {\sl Branes, central charges and
U-duality invariant BPS conditions}, hep-th/9706097.

\bibitem{bs1} J. Rahmfeld, {\sl Extremal black holes as bound states,}
Phys. Lett. {\bf B372} (1996) 198: hep-th/9512089.

\bibitem{bs2} N. Khviengia, Z. Khviengia, H. L\"u and C.N. Pope, 
{\sl Intersecting M-branes and 
bound states,}  Phys. Lett. {\bf B388} (1996) 21: hep-th/9605077.

\bibitem{bs3} M.J. Duff and J. Rahmfeld, {\sl Bound states of black holes
and other $p$-branes,} Nucl. Phys. {\bf B481} (1996) 332: hep-th/9605085.

\bibitem{dks} S. Deser, J.H. Kay and K.S. Stelle, {\sl
Renormalizability properties of supergravity,} Phys. Rev. Lett. {\bf
38} (1977) 527.

\bibitem{dk} S. Deser and J.H. Kay, {\sl Three loop counterterms for
extended supergravity,} Phys. Lett. {\bf 76B} (1978) 400.

\bibitem{kallosh81} R.E. Kallosh, {\sl Counterterms in extended
supergravities,} Phys. Lett. {\bf 99B} (1981) 122.

\bibitem{hl} P.S. Howe and U. Lindstr\"om, {\sl Higher order invariants in
extended supergravity,} Nucl. Phys. {\bf B181} (1981) 487.

\bibitem{hstsa} P.S. Howe, K.S. Stelle and P.K. Townsend, {\sl
Superactions,} Nucl. Phys. {\bf B191} (1981) 445.

\bibitem{hstsc} P.S. Howe, K.S. Stelle and P.K. Townsend, {\sl
Supercurrents,} Nucl. Phys. {\bf B192} (1981) 332.

\bibitem{hstmir} P.S. Howe, K.S. Stelle and P.K. Townsend, {\sl
Miraculous ultraviolet cancellations in supersymmetry made manifest,}
Nucl. Phys. {\bf B236} (1984) 125.

\bibitem{gdvz} M.T. Grisaru, A.E.M. van de Ven and D. Zanon, {\sl Four
loop divergences for the N=1 supersymmetric non-linear sigma model in
two dimensions,} Nucl. Phys. {\bf B277} (1986) 409.

\bibitem{kallosh92} R.E. Kallosh, {\sl Supersymmetry and black holes,}
in {\it Proc. SUSY 93 Conf., Boston, Mass.}: hep-th/9306095.

\bibitem{hawkel} S.W. Hawking and G.F.R. Ellis, {\it The Large-Scale
Structure of Space Time} (Cambridge Univ. Press, 1973).

\bibitem{gs82} S.J. Gates and W. Siegel, {\sl Linearized N=2
superfield supergravity,} Nucl. Phys. {\bf B195} (1982) 39.

\bibitem{dewit80} B. de Wit, {\sl Formulations of $N=2$ supergravity 
theories,} in {\it Proc. Europhysics Study Conf., Erice 1980}.

\bibitem{tseytlinreview} A.A. Tseytlin, {\sl Exact solutions of closed
string theory,} Class. Quantum Grav. {\bf 12} (1995) 2365: hep-th/9505052.

\bibitem{dyonic} J.M. Izquierdo, N.D. Lambert, G. Papadopoulos and
P.K. Townsend, {\sl Dyonic membranes,} Nucl. Phys. {\bf B460} (1996)
560: hep-th/9508177.

\end{thebibliography}
\end{document}